\documentstyle[12pt]{article}

\def\F{{\cal F}}
\def\A{{\cal A}}
\def\N{{\cal N}}
\def\D{\partial}
\def\L{\bar{\Lambda}}
\def\sL{\Lambda}
\def\ab{\bar a}
\def\akb{\bar a_k}

\def\bk2j{\beta_k^{(j)}}
\def\bktwo{\beta_k^{(1)}(a)}
\def\bkfour{\beta_k^{(2)}(a)}
\def\bksix{\beta_k^{(3)}(a)}
\def\bkeight{\beta_k^{(4)}(a)}

\def\rDzero{\Delta^{(0)}}
\def\rDtwo{\Delta^{(1)}}
\def\rDfour{\Delta^{(2)}}
\def\rDsix{\Delta^{(3)}}
\def\rDeight{\Delta^{(4)}}
\def\rDten{\Delta^{(5)}}
\def\rDtwelve{\Delta^{(6)}}

\def\onesum{\sum_{k=1}^{r}}
\def\twosum{\sum_{k,m=1}^{r}}
\def\threesum{\sum_{k,m,n=1}^{r}}
\def\foursum{\sum_{k,m,n,l=1}^{r}}
\def\fivesum{\sum_{k,m,n,l,o=1}^{r}}

\def\twoden{\D a_k \D a_m }
\def\threeden{\D a_k \D a_m \D a_n}
\def\fourden{\D a_k \D a_m \D a_n \D a_l}
\def\fiveden{\D a_k \D a_m \D a_n \D a_l \D a_o}
\def\instone{\F^{(1)} }
\def\insttwo{\F^{(2)} }
\def\instthree{\F^{(3)} }
\def\instfour{\F^{(4)} }
\def\instfive{\F^{(5)} }
\def\instsix{\F^{(6)} }

\begin{document} 

\begin{titlepage} 

%\begin{flushleft}
%DRAFT
%\end{flushleft}

\begin{flushright}
UCLA/99/22 \\
June, 1999
\end{flushright}
\bigskip

\begin {center}

{\LARGE Instanton Recursion Relations for the Effective \\
\medskip
Prepotential in $\N$=2 Super Yang-Mills}
\footnote{Research supported in part by the National Science
Foundation under grants PHY-9819686 and PHY-95-31023}
\\ [15mm]

Gordon Chan\footnote{E-mail:  {\tt morgoth@physics.ucla.edu}}
, 
Eric D'Hoker\footnote{E-mail:  {\tt dhoker@physics.ucla.edu}} \\ [12mm]

{\it Department of Physics and Astronomy \\ 
\medskip University of California \\ 
\medskip Los Angeles, CA 90095, USA} \\ [15mm]

\end{center}

\begin{abstract}

Linear recursion relations for the instanton corrections to the 
effective prepotential of $\N$=2 supersymmetric gauge theories with an
arbitrary number of hypermultiplets in the fundamental representation
of an arbitrary classical gauge group are derived.  The construction 
proceeds from the Seiberg-Witten solutions and the 
renormalization group type equations for the prepotential.  
Successive iterations of these recursion relations allow us to simply
obtain instanton corrections to arbitrarily high order, which we
exhibit explicitly up to 6-th order.  For gauge groups $SU(2)$ and
$SU(3)$, our results agree with previous ones.
\end{abstract}

\end{titlepage} 

\setcounter{section}{0}
\setcounter{subsection}{0}
\setcounter{equation}{0}
\setcounter{footnote}{0}

% introduction
%
\addtocounter{section}{1}
{\large {\bf \thesection. Introduction}}
\medskip
 
Over the past few years there has been great progress in understanding
non-perturbative dynamics of $\N$=2 SUSY gauge theories starting with 
the SU(2) case \cite{SW1}, further generalizations to other gauge groups
\cite{gauge} and additions of matter hypermultiplets \cite{matter}.
Non-perturbative corrections in weak coupling correspond to instanton
effects \cite{Seiberg88} which were evaluated using field theory 
techniques to one instanton 
\cite{Seiberg88}\cite{ShiftVain}\cite{Poul}\cite{Trav} 
and two instanton \cite{Dor} orders.  

Some of the previous instanton calculations using the Seiberg-Witten ansatz
were performed by solving the Picard-Fuchs equations for the 
period integrals corresponding to the quantum moduli parameters representing
the set of vacuum expectation values of the Higgs fields 
\cite{Lerche}\cite{Ohta}\cite{Ewen}.
Other previous calculations involved solving the period integrals directly
\cite{DPKi}, and were found to be in agreement with the Picard-Fuchs and
field theory results.
In an intriguing paper \cite{Matone}, a recursion relation for the 
instanton corrections to the effective prepotential $\F$ was found for 
the pure $SU(2)$ case which led us to seek a generalization of this result 
for any gauge group and number of matter hypermultiplets.
In a related development, the Seiberg-Witten equations were viewed 
analogously to the Whitham hierarchy equations \cite{Whitham}
and the WDVV equations \cite{WDVV}.  Nonlinear recursion relations for
the instanton corrections involving Jacobi $\theta$-functions (which 
themselves involve $\tau_{ij}$ as in \cite{DPKr}) were derived starting
from the Whitham hierarchy equations \cite{SpanW}. The beta function of the 
prepotential $\F$, first observed in \cite{Sonnen} and later proved in
\cite{DPKr}, is the starting point of our discussion and provides a very
direct way at calculating instanton corrections to the prepotential
$\F$ without having to perform complicated hyperelliptic integrals and
immediately obtaining rational expressions without $\theta$-functions.

The Seiberg-Witten solutions for classical gauge groups $SU(N)$,
$SO(N)$, and $Sp(N)$ and the 
renormalization group like equation for the prepotential $\F$,
led us to the discovery of a general recursion relation expressing
the $n-th$ order instanton correction to the prepotential $\F$
in terms of the $(n-1)th, \cdots , 1st$ order instanton corrections for super
Yang-Mills theory with matter hypermultiplets in the fundamental 
representation of classical gauge groups $G$.

We start off by reviewing the Seiberg-Witten solution for $\N$=2 super
Yang-Mills theory with hypermultiplets in the fundamental 
representation of any arbitrary classical gauge group.  The renormalization
group type equation for the prepotential $\F$ is discussed next and it is
shown how it can be used to determine the instanton corrections to the
prepotential to arbitrary order in an efficient manner.  Recursion 
relations for the instanton corrections are then derived and shown 
to reproduce previous results.

%
% review of results from D'Hoker, Phong, Kriechiever paper
%
\vspace{7mm}
\addtocounter{section}{1}
{\large {\bf \thesection.  The Seiberg-Witten Solution for Arbitrary
Classical Gauge Group $G$ }}
\medskip

The Seiberg-Witten (SW) ansatz gives a prescription for determining the
prepotential of the effective action for $\N$=2 supersymmetric Yang-Mills
gauge theories, as well as for determining the spectrum of BPS states.

We consider, $\N$=2 SUSY gauge theories with classical gauge 
groups $SU(r+1)$, $SO(2r+1)$, $Sp(2r)$ and $SO(2r)$, of rank r 
and number of colours $N_c$ = $r+1$, $2r+1$, $2r$, and $2r$
respectively.  We include $N_f$ hypermultiplets in 
the fundamental representation of the gauge group, with bare masses 
$m_j, \; j = 1, \cdots, N_f$.  We restrict to the asymptotically free
theories; this limits the hypermultiplet contents $N_f$.  
(ie. $N_f < 2N_c$ for $SU(N_c)$).
The classical vacuum expectation value of the gauge scalar $\phi$
is parameterized by complex moduli $\akb, k=1, \cdots, r$ as follows

\begin{equation}
\begin{array}{rlr}
SU(r+1) & \phi  =  diag[\ab_1, \cdots, \ab_r, \ab_{r+1}] &
\ab_1 + \cdots + \ab_r + \ab_{r+1} = 0 \\
SO(2r+1) & \phi  =  diag[\A_1, \cdots, \A_r, 0] & \\
Sp(2r) & \phi  =  diag[\ab_1, -\ab_1, \cdots, \ab_r, -\ab_r] & \\
SO(2r) & \phi  =  diag[\A_1, \cdots, \A_r] &
\A_k = 
\left({\begin{array}{cc} 0 & \ab_k \\ -\ab_k & 0 \end{array}}\right)
\end{array} 
\end{equation}
For generic $\akb$'s, the gauge symmetry is broken to $U(1)^r$ and
the dynamics is that of an Abelian Coulomb phase.  The
Wilson effective Lagrangian of the quantum theory to leading order
in the low momentum expansion in the Abelian Coulomb phase is
completely characterized by a complex analytic prepotential $\F(a)$.

The SW ansatz for determining the full 
prepotential $\F$ is based on a choice of a fibration of spectral
curves over the space of vacua, and of a meromorphic 1-form
$d\lambda$ on each of these curves.  The renormalized order
parameters $a_k$ of the theory, their duals $a_{D,k}$,
and the prepotential $\F$ are given by
\begin{eqnarray}
2\pi i a_k = \oint_{A_k} d\lambda, &
\; 2\pi i a_{D,k} = {\displaystyle \oint_{B_k}} d\lambda, & 
\; a_{D,k} = \frac{\D \F}{\D a_k} \label{eq:swanst}
\end{eqnarray}
with $A_k, B_k$ a suitable set of homology cycles on the 
spectral curves \cite{gauge}.

For all $\N$=2 supersymmetric gauge theories based on classical gauge
groups with $N_f$ hypermultiplets in the fundamental representation of
the gauge group, the spectral curves and meromorphic 1-forms are
\begin{eqnarray}
y^2 = A^2(x) - B(x) \nonumber \\
d\lambda = \frac{x}{y}
\left(A' - \frac{A B'}{2B}\right)dx \label{eq:rsurf} 
\end{eqnarray}
where 
\begin{eqnarray}
\begin{array}{rll}
SU(r+1) \; \; & {\displaystyle A(x)=\prod_{k=1}^{r+1}(x-\akb) } & 
\; {\displaystyle B(x)=\L^2 \prod_{j=1}^{N_f}(x+m_j) } 
\end{array} \nonumber \\
\left. \begin{array}{r} 
SO(2r+1) 
\\  Sp(2r)\footnotemark
\\ SO(2r)
\end{array} \right\} \; \;
\begin{array}{ll}
{\displaystyle A(x)=x^a \prod_{k=1}^r (x^2-\akb^2) } &
\; {\displaystyle B(x)=\L^2 x^b \prod_{j=1}^{N_f}(x^2-m_j^2) }
\end{array} \label{eq:gcurves}
\end{eqnarray}
\footnotetext{For simplicity, we restrict attention here to the
Sp(2r) case with at least two massless hypermultiplets.  The cases
with one or no massless hypermultiplets may be treated accordingly
\cite{DPKg}.}
with $\L \equiv \Lambda^q$
\begin{equation}
\begin{array}{rlll}
SU(r+1) & \; \; \; q = r+1-N_f/2 & & \\
SO(2r+1) & \; \; \; q = 2r-1-N_f & \; \; a=0  & \; \; b=2 \\
Sp(2r) & \; \; \; q = 2r+2-N_f & \; \; a=2 & \; \; b=0 \\
SO(2r) & \; \; \; q = 2r-2-N_f & \; \; a=0 & \; \; b=4 \\
\end{array} \label{eq:qnum}
\end{equation}
respectively.  The spectral curves (\ref{eq:gcurves})
for $SO(2r+1)$, $Sp(2r)$ 
and $SO(2r)$ can be obtained from the $SU(2r)$ spectral curve
by a suitable restriction on the classical moduli $\akb$'s
and masses \cite{DPKg}.

For gauge theories with classical gauge groups and asymptotically free
coupling obeying the constraint $q > 0$, general
arguments based on the holomorphicity of $\F$, perturbative 
non-renormalization theorems beyond 1-loop order, the nature of instanton
corrections, and restrictions of $U(1)_R$ invariance, constrain $\F$
to have the form
\begin{eqnarray}
\F(a) & = & \frac{2q}{\pi i} \sum_{i=1}^r a_i^2 + \frac{i}{4\pi}
\left[\sum_{\alpha} (\alpha\cdot a)^2 log \frac{(\alpha\cdot a)^2}{\Lambda^2} \right. \nonumber \\ 
&& \left. - \sum_i \sum_{j=1}^{N_f} (\lambda_i \cdot a + m_j)^2 log \frac{(\lambda_i \cdot a + m_j)^2}{\Lambda^2} \right] \nonumber \\
& + & \sum_{m=1}^{\infty} 
\frac{\Lambda^{2mq}}{2m \pi i} \F^{(m)}(a)
\label{eq:prepot}
\end{eqnarray}
where $\lambda_i =\pm e_i$ for SO and Sp and 
$\lambda_i = e_i$ for SU in an orthonormal basis $\vec{e}_i$, 
and $\alpha$ are the roots of the gauge group $G$.  
(The $SU(r)$ solution requires an additional overall 
factor of $\frac{1}{2}$).

The terms on the right side are respectively the classical
prepotential, the contribution of perturbative one-loop effects, and
$m$-instanton processes contributions \footnote{The normalization of 
the instanton contributions in the present paper differs from that 
of \cite{DPKi}\cite{DPKg} by a factor $\frac{1}{4m\pi i}$ 
for $SU(N_c)$ and $\frac{1}{2m\pi i}$ for $SO(2r+1)$,
$Sp(2r)$, and $SO(2r)$.  For our purposes, it will be convenient to
use the normalization of (\ref{eq:prepot}).}.  
$\Lambda$ is the dynamically generated scale of the theory.  

\vspace{7mm}
\addtocounter{section}{1}
{\large {\bf \thesection. Renormalization Group Type Equations}}
\medskip

In \cite{DPKr}, a renormalization group type equation for the
prepotential $\F$ was derived using the SW ansatz
equations (\ref{eq:swanst}) 
\begin{equation}
\Lambda{\D\F\over\D\Lambda} = 
\frac{q}{\pi i}
\sum_{k=1}^r \akb^2 \label{eq:renormg}
\end{equation}
up to an additive term independent of $a_k$ and $\akb$ which is
physically immaterial.  (The $SU(r)$ case
requires an additional factor of $\frac{1}{2}$).  

In \cite{DPKi} an efficient algorithm was presented for 
calculating the renormalized
order parameters $a_k$ and their duals $a_{D,k}$ in terms of
the classical order parameters $\akb$ to any order
of perturbation theory in a regime where $\L$ is small
and the $\akb$'s are well-separated.
The calculation of $a_k$ starts
off from equations (\ref{eq:swanst}) and (\ref{eq:rsurf})
producing a final result 
\begin{equation}
a_k = \sum_{m=0}^{\infty}\L^{2m}{\Delta_k^{(m)}}(\bar a)
\label{eq:ak}
\end{equation}
where we set $\Delta_k^{(0)}(\ab) \equiv \ab_k$, and we have
\begin{eqnarray}
& {\displaystyle 
\Delta_k^{(m)}(\ab) = \frac{1}{2^{2m}(m!)^2} 
\left({\D\over{\D \akb}}\right)^{2m-1}  
S_k(\akb, \ab)^m,} & m \neq 0  \label{eq:del} 
\end{eqnarray}
with
\begin{equation}
\begin{array}{rl}
SU(r+1) & \; \; S_k (x, a) = {\displaystyle
\frac{\prod_{j=1}^{N_f}(x+m_j)}{\prod_{l \neq k}(x-a_l)^2} } \\
SO(2r+1) & \; \; S_k (x, a) = {\displaystyle
\frac{x^2 \prod_{j=1}^{N_f}(x^2-m_j^2)}{(x+a_k)^2 \prod_{l \neq k}(x^2-a_l^2)^2} } \\
Sp(2r)\footnotemark[1] & \; \; S_k (x, a) = {\displaystyle
\frac{\prod_{j=1}^{N_f-2}(x^2-m_j^2)}{(x+a_k)^2 \prod_{l \neq k}(x^2-a_l^2)^2} } \\
SO(2r) & \; \; S_k (x, a) = {\displaystyle
\frac{x^4 \prod_{j=1}^{N_f}(x^2-m_j^2)}{(x+a_k)^2 \prod_{l \neq k}(x^2-a_l^2)^2} } \\
\end{array} \label{eq:sfunc}
\end{equation}
and $\L$ defined as previously (\ref{eq:qnum}).

Equations (\ref{eq:renormg})(\ref{eq:ak})(\ref{eq:del})
(\ref{eq:sfunc}) suffices
to determine the prepotential $\F$ in 
terms of the renormalized order parameters $a_k$ order by order 
in powers of $\L^2$.

%
% summary of inversion calculation
%
\vspace{7mm}
\addtocounter{section}{1}
{\large {\bf \thesection. Recursion Relation for the Prepotential $\F$}}
\medskip
	
A very direct way of deriving the form of the instanton corrections to
the prepotential $\F$ starts off from the beta function on the 
right hand side of (\ref{eq:renormg}).  Substituting the ansatz for
the prepotential (\ref{eq:prepot}) 
into the beta function (\ref{eq:renormg}), one obtains
\begin{equation}
\onesum \akb^2 = \onesum a_k^2 + 
\sum_{m=1}^{\infty} \L^{2m} \F^{(m)}(a) \label{eq:renorminst}
\end{equation}
Substituting (\ref{eq:ak}) into 
(\ref{eq:renorminst}), one obtains
\begin{eqnarray}
0 & = &  
\onesum
\left[\sum_{m=0}^{\infty}\L^{2m}\Delta_k^{(m)}(\ab) \right]^2 
- \onesum(\Delta_k^{(0)}(\ab))^2
\\ \nonumber
&&  + \sum_{m=1}^{\infty} \L^{2m} \F^{(m)}
\left(\sum_{n=0}^{\infty}\L^{2n}\Delta_k^{(n)}(\ab) \right)
\label{eq:instgen}
\end{eqnarray}
Expanding in powers of $\L^2$ in the 
last term and replacing the $\ab_k$'s with $a_k$'s, 
the $m$-th order instanton correction to the prepotential $\F$ takes on 
the form
\begin{eqnarray}
-\F^{(m)}(a) & = & \onesum
\left[\sum_{\stackrel{\scriptstyle i,j=0}{i+j=m} }^{m}\Delta_k^{(i)}(a)
\Delta_k^{(j)}(a) \right] \nonumber \\
& + & \sum_{n=1}^{m-1} \frac{1}{n!}
\sum_{\stackrel{\scriptstyle \beta_1, \cdots, \beta_{n+1} = 1}
{\beta_1 + \cdots + \beta_{n+1}=m}}^{n-1}
\sum_{\alpha_1, \cdots, \alpha_n = 1}^r
\left[ \prod_{i=1}^{n} \Delta_{\alpha_i}^{(\beta_i)}(a) \right]
\left( \prod_{j=1}^{n} \frac{\D}{\D a_{\alpha_j} } \right)
\F^{(\beta_{n+1})}(a) \nonumber \label{eq:minst}\\
\end{eqnarray}
which is a linear recursion relation for $\F^{(m)}(a)$ in terms of the lower
order instanton corrections $\F^{(m-1)}(a), \ldots, \F^{(1)}(a)$.

The intriguing part about the recursion relation (\ref{eq:minst}) 
for $\F^{(n)}(a)$ is that it is linear in 
$\F^{(n-1)}(a), \cdots, \F^{(1)}(a)$ and is valid for all classical
gauge groups with the number of hypermultiplets in the fundamental 
representation constrained by $q>0$.  Previous recursion relations
\cite{Matone} were only valid for $SU(2)$ with no hypermultiplets
and were non-linear.

\vspace{7mm}
\addtocounter{section}{1}
{\large {\bf \thesection. Instanton Expansion of the Prepotential $\F$}}
\medskip

Order by order in powers of $\L^2$, the first six instanton corrections
(\ref{eq:minst}) to the prepotential $\F$ are

\begin{eqnarray}
-\F^{(1)}(a) & = & \onesum 2\rDzero_k(a) \rDtwo_k(a) \label{eq:i2} \\
-\F^{(2)}(a) & = & \onesum
\left[ 2\rDzero_k(a) \rDfour_k(a) + (\rDtwo_k(a))^2 \right] 
+ \onesum \rDtwo_k(a) {{\D \F^{(1)}(a)}\over{\D a_k}} \\
-\F^{(3)}(a) & = & \sum_{k=1}^r
\left[ 2\rDzero_k(a) \rDsix_k(a) + 2\rDtwo_k(a)\rDfour_k(a) \right] \nonumber \\
& + & \onesum 
\left[ \rDtwo_k(a) {{\D \F^{(2)}(a)}\over{\D a_k}} \right. 
\left. + \rDfour_k(a) {{\D \F^{(1)}(a)}\over{\D a_k}} \right] \nonumber \\
& + & {1\over{2!}}\twosum
\rDtwo_k(a) \rDtwo_m(a) {{\D^2 \F^{(1)}(a)}\over{\twoden}} \\
-\F^{(4)}(a) & = & \onesum
\left[ 2\rDzero_k(a) \rDeight_k(a) + 2\rDtwo_k(a)\rDsix_k(a) + (\rDfour_k(a))^2 \right] 
\nonumber \\
& + & \onesum 
\left[ \rDtwo_k(a) {{\D \F^{(3)}(a)}\over{\D a_k}} \right. 
\left. + \rDfour_k(a) {{\D \F^{(2)}(a)}\over{\D a_k}} \right.
\left. + \rDsix_k(a) {{\D \F^{(1)}(a)}\over{\D a_k}} \right] \nonumber \\ 
& + & {1\over{2!}}\twosum
\left[ \rDtwo_k(a) \rDtwo_m(a) {{\D^2 \F^{(2)}(a)}\over{\twoden}} \right. 
+
\left. 2\rDtwo_k(a) \rDfour_m(a) {{\D^2 \F^{(1)}(a)}\over{\twoden}} \right] 
\nonumber \\
& + & {1\over{3!}}\threesum
\rDtwo_k(a) \rDtwo_m(a) \rDtwo_n(a) 
{{\D^3 \F^{(1)}(a)}\over{\threeden}} \label{eq:i8}\\
-\F^{(5)}(a) & = & \onesum
\left[ 2\rDzero_k(a) \rDten_k(a) + 2\rDtwo_k(a)\rDeight_k(a) + \right.
\left. 2\rDfour_k(a)\rDsix_k(a) \right] \nonumber \\
& + & \onesum
\left[ \rDtwo_k(a) {{\D \F^{(4)}(a)}\over{\D a_k}} \right. 
+ \rDfour_k(a) {{\D \F^{(3)}(a)}\over{\D a_k}} 
+ \rDsix_k(a) {{\D \F^{(2)}(a)}\over{\D a_k}} \nonumber \\
& + & \left. \rDeight_k(a) {{\D \F^{(1)}(a)}\over{\D a_k}} \right] 
+ {1\over{2!}}\twosum
\left[ \rDtwo_k(a) \rDtwo_m(a) {{\D^2 \F^{(3)}(a)}\over{\twoden}} \right. \nonumber \\
& + & 2\rDtwo_k(a) \rDsix_m(a) {{\D^2 \F^{(1)}(a)}\over{\twoden}} \nonumber \\
& + & \rDfour_k(a) \rDfour_m(a) {{\D^2 \F^{(1)}(a)}\over{\twoden}} 
+ \left. 2\rDtwo_k(a) \rDfour_m(a) {{\D^2 \F^{(2)}(a)}\over{\twoden}} \right]
\nonumber \\ 
& + & {1\over{3!}}\threesum
\left[ \rDtwo_k(a) \rDtwo_m(a) \rDtwo_n(a) {{\D^3 \F^{(2)}(a)}\over{\threeden}} \right.
\nonumber \\
& + & \left. 3\rDfour_k(a)\rDtwo_m(a)\rDtwo_n(a) {{\D^3 \F^{(1)}(a)}\over{\threeden}} \right]
\nonumber \\
& + & {1\over{4!}}\foursum
\rDtwo_k(a) \rDtwo_m(a) \rDtwo_n(a) \rDtwo_l(a) {{\D^4 \F^{(2)}(a)}\over{\fourden}} \\ 
-\F^{(6)}(a) & = & \onesum
\left[ 2\rDzero_k(a) \rDtwelve_k(a) + 2\rDtwo_k(a)\rDten_k(a) \right.
+ 2\rDfour_k(a)\rDeight_k(a) \nonumber \\
& + & \left. (\rDsix_k(a))^2 \right]
+ \onesum
\left[ \rDtwo_k(a) {{\D \F^{(5)}(a)}\over{\D a_k}} \right. 
+ \rDfour_k(a) {{\D \F^{(4)}(a)}\over{\D a_k}} \nonumber \\
& + & \rDsix_k(a) {{\D \F^{(3)}(a)}\over{\D a_k}} 
+ \rDeight_k(a) {{\D \F^{(2)}(a)}\over{\D a_k}}
+ \left. \rDten_k(a) {{\D \F^{(1)}(a)}\over{\D a_k}} \right] \nonumber \\
& + & {1\over{2!}}\twosum
\left[ \rDtwo_k(a) \rDtwo_m(a) {{\D^2 \F^{(4)}(a)}\over{\twoden}} \right.
+ 2\rDtwo_k(a) \rDeight_m(a) {{\D^2 \F^{(1)}(a)}\over{\twoden}} \nonumber \\
& + & \rDfour_k(a) \rDfour_m(a) {{\D^2 \F^{(2)}(a)}\over{\twoden}} 
+ 2\rDtwo_k(a) \rDfour_m(a) {{\D^2 \F^{(3)}(a)}\over{\twoden}} \nonumber \\
& + & 2\rDtwo_k(a) \rDsix_m(a) {{\D^2 \F^{(2)}(a)}\over{\twoden}} 
+ \left. 2\rDfour_k(a) \rDsix_m(a) {{\D^2 \F^{(1)}(a)}\over{\twoden}} \right]
\nonumber \\ 
& + & {1\over{3!}}\threesum
\left[ \rDtwo_k(a) \rDtwo_m(a) \rDtwo_n(a) {{\D^3 \F^{(3)}(a)}\over{\threeden}} \right. \nonumber \\
& + & 3\rDtwo_k(a) \rDtwo_m(a) \rDsix_n(a) {{\D^3 \F^{(1)}(a)}\over{\threeden}} \nonumber \\
& + & 3\rDtwo_k(a) \rDtwo_m(a) \rDfour_n(a) {{\D^3 \F^{(2)}(a)}\over{\threeden}} \nonumber \\
& + & \left. 3\rDtwo_k(a)\rDfour_m(a)\rDfour_n(a) {{\D^3 \F^{(1)}(a)}\over{\threeden}} \right]
\nonumber \\
& + & {1\over{4!}}\foursum
\left[ \rDtwo_k(a) \rDtwo_m(a) \rDtwo_n(a) \rDtwo_l(a) {{\D^4 \F^{(2)}(a)}\over{\fourden}} \right. \nonumber \\
& + & \left. 4\rDfour_k(a) \rDtwo_m(a) \rDtwo_n(a) \rDtwo_l(a) {{\D^4 \F^{(1)}(a)}\over{\fourden}} \right] \nonumber \\
& + & {1\over{5!}}\fivesum
\rDtwo_k(a) \rDtwo_m(a) \rDtwo_n(a) \rDtwo_l(a) \rDtwo_o(a) {{\D^5 \F^{(1)}(a)}\over{\fiveden}} \nonumber \\
\end{eqnarray}

A closer examination of the recursion relation 
(\ref{eq:minst}) for the prepotential $\F$ reveals
that there is always a term of the form 
\begin{equation}
2\onesum \rDzero_k(a) \Delta_k^{(n)}(a) = 
2\onesum a_k \Delta_k^{(n)}(a) \label{eq:funnyfirst}
\end{equation}
When performing explicit calculations for special cases of $N_c$ and $N_f$,
it is useful to rewrite terms of the form (\ref{eq:funnyfirst})
so that there are no $a_k$'s sitting out in front.  Using the
definition (\ref{eq:sfunc}) of $S_k(x, a)$ and performing contour
integrals in the complex plane by residue methods as in \cite{DPKi},
it can be shown that
\begin{equation}
2\onesum a_k \Delta_k^{(n)}(a) = -\frac{(2n-1)}{2^{2n-1} (n!)^2}
\onesum
\left(\frac{\D}{\D a_k}\right)^{2n-2} S_k(a_k, a)^n 
\end{equation}
up to an $a_k$ independent term that is physically immaterial for $q>0$.

\vspace{7mm}
\setcounter{subsection}{0}
\setcounter{subsubsection}{0}
\addtocounter{section}{1}
{\large {\bf \thesection. Comparison with Previous Results}}
\medskip

In order to make explicit comparisons with results in the literature,
the instanton corrections have to be rewritten in terms of
symmetric polynomials in the $a_k$'s as follows.

For $SU(2)$, the existing results in the literature have the instanton
expressions expressed in terms of
\begin{eqnarray}
a_1 & = & 2a \nonumber \\ 
a_2 & = & -2a 
\end{eqnarray}

Solving the recursion relation (\ref{eq:minst}) for the pure $SU(2)$ case, 
the explicit form for the n-th order instanton correction to the 
prepotential $\F$ was determined to be

\begin{eqnarray}
\F^{(n)}(a) =  \frac{1}{(2a)^{4n-2} } 
\sum_{j=1}^{n} \left({\begin{array}{c} 4n-3 \\ j-1 \end{array}} \right) \frac{(-1)^{j-1}}{j} \sum_{\stackrel{\scriptstyle n_1, \cdots, n_j =1}{n_1+ \cdots + n_j = n}}^{n}b_{n_1}\cdots b_{n_j} \nonumber \\
\end{eqnarray}
where
\begin{equation}
b_n = \frac{(2n-3)!!}{(n!)^2} 
\end{equation}
which agrees with previous results \cite{Lerche}\cite{DPKi}
\footnote{Our results agree exactly with those of \cite{Lerche} 
to eight instantons with the replacement 
$\Lambda^2 \rightarrow \frac{\Lambda^2}{2}$.}.

Explicit evaluations for $N_f = 0,1,2,3$ were performed, with $N_f=3$
summarized here.
\begin{eqnarray*}
\instone & = & \frac{1}{2^2 a^2}
\left[ a^2(m_1+m_2+m_3) + m_1 m_2 m_3 \right] \\
\insttwo & = & \frac{1}{2^8 a^6}
\left[ a^6 + a^4(m_1^2+m_2^2+m_3^2)- a^2(m_1^2 m_2^2 +m_1^2 m_3^2 +m_2^2 m_3^2) + 5m_1^2 m_2^2 m_3^2 \right] \\
\instthree & = & \frac{m_1 m_2 m_3}{2^{11} a^{10}}
\left[ -3a^6 + 5a^4 (m_1^2 + m_2^2 + m_3^2) - 7a^2 (m_1^2 m_2^2 + m_1^2 m_3^2 + m_2^2 m_3^2) \right. \\
&+ & \left. 9m_1^2 m_2^2 m_3^2 \right] \\
\instfour & = & \frac{1}{2^{20} a^{14}}
\left\{ a^{12} - 6a^{10}(m_1^2 + m_2^2 + m_3^2) + a^8 [5(m_1^4 +m_2^4 +m_3^4) \right. \\
& + & 100(m_1^2 m_2^2 + m_1^2 m_3^2 + m_2^2 m_3^2)] + a^6 [1176 m_1^2 m_2^2 m_3^2 \\
& - & 126(m_1^4 m_2^2 + m_1^2 m_2^4 + m_1^4 m_3^2 + m_1^2 m_3^4 + m_2^4 m_3^2 + m_2^2 m_3^4)] \\
& + & a^4 [153(m_1^4 m_2^4 + m_1^4 m_3^4 + m_2^4 m_3^4)+ 1332m_1^2 m_2^2 m_3^2(m_1^2 + m_2^2 + m_3^2)] \\
& - & 1430a^2 m_1^2 m_2^2 m_3^2 (m_1^2 m_2^2+ m_1^2 m_3^2 + m_2^2 m_3^2) 
\left. + 1469 m_1^4 m_2^4 m_3^4 \right\} \\
\instfive & = & \frac{m_1 m_2 m_3}{2^{23} a^{18}}
\left\{35 a^{12} -210a^{10} (m_1^2 + m_2^2 + m_3^2) \right. \\
& + & a^8 [207(m_1^4 + m_2^4 + m_3^2) + 
1260(m_1^2 m_2^2 + m_1^2 m_3^3 + m_2^2 m_3^2)] \\
& - & 1210a^6 (m_1^4 m_2^2 + m_1^2 m_2^4 + m_1^4 m_3^2 + m_1^2 m_3^4 
+ m_2^4 m_3^2 + m_2^2 m_3^4) \\
& + & a^4[1131(m_1^4 m_2^4 + m_1^4 m_3^4 + m_2^4 m_3^4) 
+ 5960m_1^2 m_2^2 m_3^2(m_1^2 + m_2^2 + m_3^2)] \\
& - & 5250 a^2 m_1^2 m_2^2 m_3^2 (m_1^2 m_2^2 + m_1^2 m_3^2 + m_2^2 m_3^2)
\left. + 4471 m_1^4 m_2^4 m_3^4 \right\} \\
\instsix & = & \frac{1}{2^{29}a^{22} }
\left\{5a^{16}(m_1^2 + m_2^2 + m_3^2) - a^{14}[210(m_1^2 m_2^2 + m_1^2 m_3^2 + m_2^2 m_3^2) \right. \\
& + & 14(m_1^4 + m_2^4 + m_3^4)] + a^{12}[9(m_1^6 + m_2^6 + m_3^6) + 
6507 m_1^2 m_2^2 m_3^2 \\
& + & 801(m_1^4 m_2^2 + m_1^2 m_2^4 + m_1^4 m_3^2 + m_1^2 m_3^4 + m_2^4 m_3^2
+ m_2^2 m_3^4)] \\
& - & a^{10}[660(m_1^6 m_2^2 + m_1^2 m_2^6 + m_1^6 m_3^2 + m_1^2 m_3^6 
+ m_2^6 m_3^2 + m_2^2 m_3^6) \\
& + & 330(m_1^4 m_2^4 + m_1^4 m_3^4 + m_2^4 m_3^4)
+ 24420m_1^2 m_2^2 m_3^2 (m_1^2 + m_2^2 + m_3^2)] \\
& + & a^8[2769(m_1^6 m_2^4 + m_1^4 m_2^6 + m_1^6 m_3^4 + m_1^4 m_3^6 
+ m_2^6 m_3^4 + m_2^4 m_3^6) \\
& + & m_1^2 m_2^2 m_3^2 (19851(m_1^4 + m_2^4 + m_3^4) 
+ 87945(m_1^2 m_2^2 + m_1^2 m_3^2 + m_2^2 m_3^2))] \\ 
& - & a^6[295050 m_1^4 m_2^4 m_3^4 + 2310(m_1^2 m_2^2 + m_1^2 m_3^2
+ m_2^2 m_3^2) \\
& + & 69510m_1^2 m_2^2 m_3^2 (m_1^4 m_2^2 + m_1^2 m_2^4 + m_1^4 m_3^2 
+ m_1^2 m_3^4 + m_2^4 m_3^2 + m_2^2 m_3^4)] \\
& + & a^4m_1^2 m_2^2 m_3^2 [53839(m_1^4 m_2^4 + m_1^4 m_3^4 + m_2^4 m_3^4)
+224485m_1^2 m_2^2 m_3^2 (m_1^2+m_2^2+m_3^2)] \\
& - & 166896a^2 m_1^4 m_2^4 m_3^4 (m_1^2 m_2^2 +m_1^2 m_3^2 +m_2^2 m_3^2)
+ \left. 121191 m_1^6 m_2^6 m_3^6 \right\}
\end{eqnarray*}
\medskip

A check of the hypermultiplet decoupling limits of 
the $N_f=3$ instanton corrections,
by letting $\Lambda_3 m_3 = \Lambda_2^2$ and
sending $m_3 \rightarrow \infty$, reproduces the $N_f=2$ results.
A further decoupling of a second hypermultiplet, by letting 
$\Lambda_2 m_2 = \Lambda_1^2$ and sending $m_c \rightarrow \infty$,
reproduces the $N_f=1$ results.
Comparison with results in the literature \cite{Ohta}\cite{DPKi}\cite{SpanW}
show an agreement to four instantons up to a redefinition of the
$\akb$'s as discussed in \cite{DPKi}.

For $SU(3)$, the existing results in the literature have the instanton
corrections expressed in terms of the invariant $SU(3)$ symmetric 
polynomials $u,v$ and the discriminant $\Delta$
\begin{eqnarray}
u & = & - a_1 a_2 - a_1 a_3 - a_2 a_3 \nonumber \\
v & = & a_1 a_2 a_3 \nonumber \\
\Delta & = & 4u^3 - 27 v^3
\end{eqnarray}
and the $p$-th symmetric mass polynomials
\begin{equation}
t_p(m) = \sum_{j_1 < \cdots < j_p} m_{j_1}\cdots m_{j_p}
\end{equation}

Explicit evaluations for $N_f = 0,1,2,3,4,5$ were performed, and are 
summarized here for $N_f=0$.
\begin{eqnarray*}
\instone & = & \sL^6 \frac{3u}{\Delta} \\
\insttwo & = & \frac{\sL^{12} u}{16} 
\left[ \frac{10935 v^2}{\Delta^3} + \frac{153}{\Delta^2} \right] \\
\instthree & = & \frac{3\sL^{18} u}{16}
\left[ \frac{4782969 v^4}{2\Delta^5} + \frac{161109 v^2}{2\Delta^4} \right.
\left. + \frac{385}{\Delta^3} \right] \\
\instfour & = & \frac{\sL^{24} u}{4096}
\left[ \frac{1707362095023v^6}{\Delta^7} + \frac{91216001799 v^4}{\Delta^6} \right.
+ \frac{1254600981 v^2}{\Delta^5} \\
&& \left. + \frac{3048885}{\Delta^4} \right] \\
\instfive & = & \frac{5\sL^{30} u}{4096}
\left[ \frac{3788227372819653 v^8}{10\Delta^9} + \frac{277223767370307 v^6}{10 \Delta^8} \right. \\
&& +\frac{6447389599341 v^4}{10\Delta^7} + \frac{50110037721 v^2}{10\Delta^6}
\left. + \frac{7400133}{\Delta^5} \right] \\
\instsix & = & \frac{3\sL^{36}u}{65535}
\left[ \frac{24952152189682606959 v^{10} }{2\Delta^{11} } \right.
+ \frac{2319087386959542567 v^8}{2\Delta^{10} } \\
&& + \frac{38185135433846901 v^6}{\Delta^9}
+ \frac{525166021552761 v^4}{\Delta^8} \\
&& + \frac{5323867298775 v^2}{2\Delta^7}
+\left. \frac{5295230391}{2\Delta^6} \right] 
\end{eqnarray*}
A check of successive hypermultiplet decoupling limits of the 
$N_f=5$ instanton corrections reproduces 
all of the $N_f < 5$ cases accordingly.
Comparison with results in the literature 
\cite{Lerche}\cite{DPKi}\cite{Ewen}\cite{SpanW}
show an agreement to three instantons up to a redefinition of the
$\akb$'s as discussed in \cite{DPKi}.

\vspace{7mm}
\setcounter{subsection}{0}
\addtocounter{section}{1}
{\large {\bf \thesection. Summary}}
\medskip

The recursion relations discovered in this paper improve 
considerably the ability to evaluate explicitly the non-perturbative 
instanton corrections to $\N$=2 super Yang-Mills theories. 
Possible extensions to other problems like
the strongly coupled $\N$=2 SUSY $SU(N_c)$ Seiberg-Witten problem 
\cite{DPs}\cite{SpanS} were also investigated \cite{GCu}.

\vspace{7mm}
%
% acknowledgments
%
\setcounter{subsection}{0}
\addtocounter{section}{1}
{\large {\bf Acknowledgments}}
\medskip

We are grateful to D.H. Phong for several helpful discussions and 
collaboration at the very early stage of this work.
G.C. would like to thank NSERC for financial support.

\vspace{14mm}
{\LARGE{\bf Appendix}}

\appendix
\setcounter{subsection}{0}

\vspace{7mm}
\addtocounter{section}{1}
{\large {\bf \thesection.  Classical Moduli in Terms of Quantum Moduli} }
\medskip

Another way of evaluating the beta function (\ref{eq:renormg}) 
of the prepotential $\F$ involves inverting (\ref{eq:ak})
to get
\begin{equation}
\akb \equiv a_k + 
\sum_{m=1}^{\infty} \L^{2m} \beta_k^{(m)}(a) \label{eq:betas}
\end{equation}	
where the $\beta_k(a)$'s are functions of the renormalized order
parameters $a_k$. 

A very direct way of deriving the form of the $\beta_k(a)$'s
involves starting off with (\ref{eq:betas}) and
substituting in equation (\ref{eq:ak}) to get
\begin{eqnarray}
0 = \sum_{m=1}^{\infty}\L^{2m}{\Delta_i^{(m)}}(\bar a) +
\sum_{m=1}^{\infty}\L^{2m}{\beta_i^{(m)}}
\left( \sum_{m=0}^{\infty}\L^{2m}{\Delta_k^{(m)}}(\bar a) \right)
\label{eq:generate}
\end{eqnarray}
Expanding in powers of $\L^2$ in the second term 
and replacing the $\akb$'s with $a_k$'s, one obtains
\begin{eqnarray}
-\beta_k^{(m)}(a) & = & \Delta_k^{(m)}(a) \nonumber \\
& + & \sum_{n=1}^{m-1} \frac{1}{n!}
\sum_{\stackrel{\scriptstyle \beta_1, \cdots, \beta_{n+1} = 1}
{\beta_1 + \cdots + \beta_{n+1}=m}}^{n-1}
\sum_{\alpha_1, \cdots, \alpha_n = 1}^r
\left[ \prod_{i=1}^{n} \Delta_{\alpha_i}^{(\beta_i)}(a) \right]
\left( \prod_{j=1}^{n} \frac{\D}{\D a_{\alpha_j} } \right)
\beta_k^{(\beta_{n+1})}(a) \nonumber \label{eq:bk}\\
\end{eqnarray}
Order by order in powers of $\L^2$, the first few $\beta_k(a)$'s are 
\begin{eqnarray*}
-\bktwo & = & \rDtwo_k(a)  \\
-\bkfour & = & \rDfour_k(a) + \sum_{l=1}^r \rDtwo_l(a)
{{\D \bktwo}\over{\D a_l}} \\
-\bksix & = & \rDsix_k(a) + \sum_{l=1}^r
\left[ \rDtwo_l(a) {{\D \bkfour}\over{\D a_l}} \right.
\left. + \rDfour_l(a) {{\D \bktwo}\over{\D a_l}} \right] \\
& + & {1\over{2!}}\sum_{l,m=1}^r \rDtwo_l(a) \rDtwo_m(a)
{ {\D^2 \bktwo}\over {\D a_l \D a_m} } \\
-\bkeight & = & \rDeight_k(a) + \sum_{l=1}^r
\left[ \rDtwo_l(a) {{\D \bksix}\over{\D a_l}} \right.
\left. + \rDfour_l(a) {{\D \bkfour}\over{\D a_l}} \right. \\
& + & \left. \rDsix_l(a) {{\D \bktwo}\over{\D a_l}} \right]
+ {1\over{2!}}\sum_{l,m=1}^r 
\left[\rDtwo_l(a) \rDtwo_m(a) {{\D^2 \bkfour}\over{\D a_l \D a_m}} \right. \\
& + & \left. 2\rDtwo_l(a) \rDfour_m(a) {{\D^2 \bktwo}\over{\D a_l \D a_m}} \right] \\
& + & {1\over{3!}}\sum_{l,m,n=1}^r
\rDtwo_l(a) \rDtwo_m(a) \rDtwo_n(a)
{ {\D^3 \bktwo}\over{\D a_l \D a_m \D a_n} }
\end{eqnarray*}
Substituting (\ref{eq:bk}) into (\ref{eq:renorminst})
reproduces the instanton corrections to the prepotential
(\ref{eq:minst}) order by order in $\L^2$.

%
% bibliography
%

\end{document}